\documentstyle[prl,floats,psfig,aps]{revtex}

\newcommand{\be}{\begin{equation}}
\newcommand{\ee}{\end{equation}}
\newcommand{\bea}{\begin{eqnarray}}
\newcommand{\eea}{\end{eqnarray}}
\newcommand{\sch}{Schwarzschild }

\input epsf

\begin{document}

\draft


\title{Adaptive computation of gravitational waves from
black hole interactions}

\author{Philippos Papadopoulos${}^{(1)}$, Edward
Seidel${}^{(1,2,3)}$, and Lee Wild${}^{(1)}$}

\address{
${}^{(1)}$ Max-Planck-Institut f{\"u}r Gravitationsphysik,
Schlaatzweg 1, 14473 Potsdam, Germany
}
\address{
${}^{(2)}$ National Center for Supercomputing Applications,
Beckman Institute, 405 N. Mathews Ave., Urbana, IL, 61801
}
\address{
${}^{(3)}$ Departments of Astronomy and Physics,
University of Illinois, Urbana, IL 61801
}
\date{\today}
\maketitle

\begin{abstract}
We construct a class of linear partial differential equations
describing general perturbations of non-rotating black holes in 3D
Cartesian coordinates.  In contrast to the usual approach, a single
equation treats all radiative $\ell -m$ modes simultaneously, allowing
the study of wave perturbations of black holes with arbitrary 3D
structure, as would be present when studying the full set of nonlinear
Einstein equations describing a perturbed black hole.  This class of
equations forms an excellent testbed to explore the computational
issues of simulating black spacetimes using a three dimensional
adaptive mesh refinement code.  Using this code, we present results
from the first fully resolved 3D solution of the equations describing
perturbed black holes.  We discuss both fixed and adaptive
mesh refinement, refinement criteria, and the computational savings
provided by adaptive techniques in 3D for such model problems of
distorted black holes.
\end{abstract}

\vskip2pc


\section{Introduction}
\label{sec:Intro}

The spiraling coalescence of black hole binaries is considered among
the most important sources for the upcoming worldwide network of
gravitational wave observatories.  Essential for the effective
detection and interpretation of a coalescing black hole event will be
the estimation of the gravitational waveform during the
merger~\cite{Flanagan97a,Flanagan97b}, a very non-linear event, which
must be computed by direct numerical integration of the Einstein
equations.  The amount of reliable detail that can be provided by
numerical relativity simulations of such events bears directly on the
amount of astrophysical information that can be determined from the
observations themselves.  The simulations required to study such
realistic astrophysical events must be fully 3-dimensional.  However,
the computational requirements for solving the complete set of
Einstein equations in 3D are highly non-trivial.

Current 3D black hole simulations produce accurate wave-forms
for a certain integration time, yet problems with the inner and outer
boundaries, and/or poor resolution of certain features, limit the
duration of the simulations to a fraction of the physically
interesting space-time.  With the exception of very recent simulations
based on a 3D characteristic formulation of the
equations~\cite{Gomez98a}, this is typical of all 3D numerical black
hole simulations to date (see, e.g.,
~\cite{Allen97a,Allen98a,Camarda97a,Camarda97b,Anninos94c,Anninos94d}).

From the perspective of {\em gravitational wave astronomy}, the availability
of adequate resolution in those computations has twofold significance:
First, features unique to general relativity would develop in the
strong field region near the black holes during the merger, and those
effects must be quantitatively captured. This is the problem of
correctly modeling the source. Next, the imprint stamped on the
radiative part of the metric by the source motion must be given plenty
of room and resolution to grow and propagate, up until the point where
reading it off will be unambiguous. This is the problem of correctly
capturing the gravitational signal. It is obvious from the above
discussion, that despite the fact the equations describing black
holes can be purely vacuum, there are several scales involved 
(ultimately deriving from the nature of the imposed initial data).

One can attempt a simple overview of those scales and the
corresponding resolution requirements: The shape of the black hole
effective potential for gravitational waves, well known from
perturbation theory studies~\cite{Chandrasekhar83}, can be used as a
guiding principle. There are three different regions of importance.
Those can be qualitatively appreciated by inspection of
Fig.~\ref{fig:potential}. The picture shows the effective black hole
potential for generic perturbations (more details are given in
section~\ref{sec:Model}). The relevant regions span roughly three
decades in a logarithmic radial scale, from $0.1M$ to $100M$. The
inner region has features of the order of $1M$, i.e., the scale of the
horizon for each black hole. Generic strong field regions, modeled in
this picture by the potential peak, will encompass the binary and its
effective potential, i.e., a domain of roughly 10M. Finally, for a
domain extending considerably away from the holes ($\approx 100M$),
adequate resolution is required for allowing outgoing radiation to
form and propagate.

The relative balance of those regions in the resolution budget will
depend heavily on the details of the geometrical and numerical
formalism.  This has led many groups to the proposal and development
of concepts which aim (among other things) at reducing the resolution
requirements.  The inner region ($r<1M$), being causally disconnected
from the exterior, essentially asks for an effective recipe of {\em no
computation}, so that the scale that needs to be resolved is bounded
below by the size of the horizons.  Apparent horizon boundary
conditions (see, e.g., ~\cite{Seidel92a,Cook97a}) or characteristic
based evolution~\cite{Gomez98a} are the current techniques in this
direction.  The range of $1M-10M$ is thought of as the interesting
domain in which the merger drama will unfold, and hence the bulk of
the resolution should be provided here.  In the exterior region a
gradual and subtle transition to the weak field occurs.  The domain in
the range $10M-100M$ may be handled cleanly and efficiently, for
example, with a close-in Cauchy-characteristic
matching~\cite{Bishop98a}.  Other possibilities include the evolution
of perturbed spacetimes~\cite{Abrahams97a} or other exact approaches
as for example the evolution of conformally compactified equations on
hyperboloidal hypersurfaces~\cite{Huebner96}.  Even as those
approaches gradually mature, it is important to assess the
possibilities of managing the resolution requirements from within the
framework of computational science.  This paper aims to present some
first insights in this direction.

In 3D black hole coalescence simulations, which will likely be
performed in 3D Cartesian coordinates, we will need to resolve waves
with wavelengths of order $5M$ or less, where $M$ is the mass of the
black hole.  Although for Schwarzschild, the fundamental $\ell=2$
normal mode wavelength is $16.8M$, higher modes, such as $\ell=4$ and
above, have wavelengths of $8M$ and below.  More important, for very
rapidly rotating Kerr black holes, which are expected to be formed in
realistic astrophysical black hole coalescence, the modes are shifted
down to significantly shorter
wavelengths\cite{Flanagan97a,Flanagan97b}.  On the other hand, the
size of the black hole is also about $2M$.  As we need of order 20
grid zones to resolve a single wavelength, we very conservatively
estimate a required grid resolution of about $\Delta x = \Delta y =
\Delta z \approx 0.2M$.  (The best simulations of distorted or
colliding black holes in 3D, for which accurate waveforms can be
obtained, already use higher resolution, although this is partly
needed due to slicing effects\cite{Camarda97b,Allen98a}.  It is
presently unknown whether even higher resolution near the black holes
will be required for effective and general apparent horizon boundary
conditions.)  For simulations of time scales of order $t \propto
10^{2}-10^{3}M$, which will be required to follow coalescence, the
outer boundary will probably be placed at a distance of at least $R
\propto 100M$ from the coalescence, requiring a Cartesian simulation
domain of about $200M$ across.  This leads to roughly 1000 grid zones
in each dimension, or about $10^{9}$ grid zones in total.  As 3D codes
to solve the full Einstein equations have typically 100 variables to
be stored at each location, and simulations are performed in double
precision arithmetic, this leads to a memory requirement of order 1000
Gbytes!

The largest supercomputers available to scientific research
communities today have only about $\frac{1}{20}$ of this capacity, and
machines with such capacity will not be available for some years.
Furthermore, if one needs to double the resolution in each direction
for a more refined simulation, the memory requirements increase by an
order of magnitude.  Although such estimates will vary, depending on
the ultimate effectiveness of inner or outer boundary treatments,
gauge conditions, etc., they indicate that barring some unforeseen
simplification, some form of adaptive computation that places
resolution only where it is required is not only desirable, but
essential for such problems.

The subject of this paper, then, is to elucidate the potential of
adaptive methods in three-dimensional black hole simulations.
Adaptive mesh refinement (AMR), has been an important development in
the last two decades, in applied fields employing large scale
numerical computation.  It is nevertheless still a frontier area for a
wide class of problems. Integration algorithms for hyperbolic 3D
partial differential equations, using adaptive finite differences are
now increasingly explored~\cite{Leveque98a}.  The application of such
techniques to general relativistic problems has already shown great
promise in one dimensional implementations of the Berger-Oliger AMR
method~\cite{Berger84}. For example, Choptuik's fundamental work in
critical phenomena~\cite{Choptuik93} was enabled by the use of
adaptive techniques~\cite{Choptuik89}, which were later applied by
other groups to similar problems~\cite{Abrahams93a,Stewart96};
spherical black hole evolutions were carried out with AMR techniques
in~\cite{Masso94b,Wild96}.

More extensive applications of adaptive methods in higher dimensional
relativistic problems, are clearly in order. The application of such
techniques has primarily been hampered by the complexities associated
with computationally implementing the hierarchical tree structure of
nested refined sub grids~\cite{Parashar95} that underlies those
methods.  In order to circumvent such issues, Wild and
Schutz~\cite{Wild98a} proposed a simpler adaptive mesh refinement
data-structure based on {\em hierarchical linked lists}.  Preliminary
tests of this system were applied to model problems involving scalar
fields~\cite{Wild96}.

Building on this work, we adopt here the hierarchical linked
list~(HLL) approach, in its full 3D genre. The still fermenting issue
of the correct mathematical approach(es) to the black hole problem,
suggests that our exploration of 3D adaptivity in black hole
computations would be initially best served by the use of {\em model}
equations. To this end, we introduce here a class of 3D linear partial
differential equations, inspired by the theory of perturbations of
black holes, which we then use to study issues pertaining to adaptive
computations. There is strong physical motivation for those models,
coming from the comparison of perturbative studies of black hole
interactions with full non-linear computations, which has shown good
agreement~\cite{Allen97a,Price94a,Price94b,Abrahams95c,Baker96a,Gleiser96b}
for a large range of currently feasible simulations.  The linear model
equations we introduce, are completely, even if artificially,
three-dimensional, i.e., they retain no memory of the angle and time
separated equations from which they may arise. Scalar, electromagnetic
and axial gravitational perturbations of non-rotating black holes are
obtained with the appropriate choice of model parameters. We explore
aspects of the models that increase their utility as a calibration
tool for the adaptive infrastructure. It is apparent that we exploit
here the, by now well developed, interface between perturbation
theorists and numerical relativists~\cite{argument}.

We apply fully adaptive mesh refinement techniques for the first time
in a 3D relativistic calculation, modeling the dynamics of distorted
black holes.  We examine a variety of different ways of refining the
solution, which we believe are going to be major themes in future
applications of AMR to the black hole problem.  These include({\em a})
fixed refinement regions that are prescribed in advance, exploiting
prior knowledge of the regions that will require high resolution,
({\em b}) different geometries of refinement regions, and ({\em c})
fully adaptive calculations that follow certain features that develop
during the calculation.  We discuss the effects of different
refinement criteria, and hybrid refinement, which mixes predefined
refinement regions with fully adaptive refinement.  We show that these
techniques can be used very effectively to resolve complex wave
structures emitted by systems such as perturbed black holes.

The organization of the paper is as follows: In Sec.~\ref{sec:Model}
we outline a class of 3D linear partial differential equations that
model weak field black hole dynamics. We then study some aspects on
the behavior of their solutions, in particular as they introduce the
framework in which AMR will be tested. Our Adaptive Mesh Refinement
techniques are summarized in Sec.~\ref{sec:code}. In
Section~\ref{sec:static} we present details of AMR simulations that
focus on static, prescribed in advance refinement. In
Section~\ref{sec:dynamic} we present dynamical AMR simulations. We
close (Sec.~\ref{sec:summary}) with a discussion and summary of our
results.

Unless stated otherwise, distances refer to the isotropic coordinate
system and are expressed in units of the black hole mass $M$.

\section{Model equations for black hole perturbations}
\label{sec:Model}

\subsection{Linear three-dimensional PDE's}

In this section we describe a general class of three-dimensional
partial differential equations that simulate aspects of black hole
perturbations. As our standard example we start with the
Regge-Wheeler equation~\cite{Regge57}:
\begin{equation}
Z_{,r_{*}r_{*}} - Z_{,tt} = \frac{\Delta}{r^2} \left[\ell(\ell+1)r -
6M\right] Z = V(r) Z \; .
\label{eq:rw}
\end{equation}
where $r$ is the \sch radial coordinate, $\Delta=r(r-2M)$,
$r_{*}$ is a logarithmic ``tortoise'' coordinate $dr_{*}=r^2 dr /
\Delta$ that relegates the coordinate location of the event horizon to
negative infinity, $\ell$ is the spherical harmonic index, and $M$ is
the mass of the black hole. This equation describes {\em axial
perturbations}(odd-parity) of non-rotating black holes, with $Z$ being
an appropriate combination of metric perturbations and their derivatives.
Note that, because the background is spherically symmetric,
the equation is independent of the azimuthal perturbation
parameter $m$, hence all perturbations of the same $\ell$ value obey
the same equation. As we will see below, it is this particular
$\ell$ dependence  which allows one to write a single 3D
equation to study the simultaneous excitation of various angular
perturbations.

Similar equations, with different potential terms, govern the polar
(even-parity) perturbations, as well as the propagation of scalar and
electromagnetic waves in the background of a non-rotating hole.
Although the equation governing even parity perturbations, first
developed by Zerilli, has a more complicated potential, it has
identical quasi-normal mode structure, as was shown by
Chandrasekhar~\cite{Chandrasekhar83}. However, the $\ell$ dependence
of this equation is such that it is difficult or impossible to develop
from it a single 3D equation that treats the different angular modes
simultaneously.  For these reasons, in what follows we will focus on
the Regge-Wheeler and similar equations.

In the first step of the construction procedure we restore the
symmetry between spatial directions. This is achieved minimally with
the introduction of isotropic coordinates,
\begin{eqnarray}
  r & = & \bar{r} A^{2} \; , \\ \nonumber
  \bar{r} & = & \sqrt{x^2 + y^2 + z^2} \; ,
\end{eqnarray}
where $A=1+M/2\bar{r}$.

Next, we introduce the field
\begin{equation}
  \phi(t,\bar{r},\theta,\phi) = \sum_{\ell =0}^{\infty}
\sum_{m=-\ell}^{\ell}
  Z_{\ell m}(t,\bar{r}) Y_{\ell m}(\theta,\phi) \; ,
\label{superpose}
\end{equation}
which encodes the aggregate evolution of the separated functions
$Z_{\ell m}$. It is then straightforward to establish, that
the scaled field variable $\Phi = \phi/\bar{r}A^2$ satisfies an equation
of the type
\begin{equation}
\Phi_{,tt} = a_{1}({\bf\bar{x}}) \nabla^{2} \Phi
           + a_{2}({\bf\bar{x}}) \, \hat{n} \cdot \bar{\nabla} \Phi
           + a_{3}({\bf\bar{x}}) \Phi
           + a_{4} \, \rho  \;  ,
\label{eq:rw3d}
\end{equation}
where $a_{i}$ are purely radial functions, $\hat{n}$ is the unit
radial vector, $\bar{\nabla}$ is the gradient and $\nabla^{2}$ the
Laplacian operator, all in Euclidean isotropic three-space. Most
notably for our purposes, in this equation, the $\nabla^{2}$ operator
has absorbed the angular operator $L^2 \phi = - \ell (\ell+1) \phi$
and with it, all dependence on $\ell$.

The coefficient $a_{1} = B^2/A^6$ determines the coordinate speed
of waves ($B=1-M/2\bar{r}$). The coefficient $a_{2} = M^2
B/2\bar{r}^3 A^7$
governs the apparent {\em damping} behavior near the horizon.
The coefficient $a_{3}$ is the effective potential for the field
and determines among others the frequency and decay rate of
quasi-normal mode radiation. As mentioned above, perturbations of
different spin have different potentials than Eq.~(\ref{eq:rw}), the
remaining structure of the equation being the same.
Following the spirit of~\cite{Price72} we introduce a
general effective potential for a non-rotating black hole as
\begin{equation}
a_{3}  =  - \frac{B^2}{\bar{r}^3 A^8} \left( K_{1} - 2M +
\frac{K_{2}}{\bar{r} A^2} \right) \; ,
\label{eq:vterms}
\end{equation}
where $K_{1},K_{2}$ are real parameters.
The dominant contribution to this potential comes from the Riemann
curvature, whose components behave as $M/r^3$ for large $r$. Fields of
different spin have different quantitative interactions with the
curvature potential and this is encoded in the value of $K_{1}$.  For
fields of spin number $0,1,2$, i.e., scalar, electromagnetic and axial
gravitational perturbations, the coefficient $K_{1}$ has values
$2M,0,-6M$ respectively. Curvature effects are finite at the horizon
which implies that they are redshifted to zero in a static \sch time
slicing (the factor $B^2$ in~(\ref{eq:vterms})).  Higher order
contributions to the curvature are modeled by the coefficient
$K_{2}$, which represents a ``deformation'' of the potential, e.g.,
for a charged black hole $K_{2} = 4 e^2$.  A massive scalar field of
mass $m$ would have an additional potential term given by $a_{3} = m^2
B^2/A^2$, whereas excitations by a source are generically modeled
with the term $a_{4} \, \rho(t,{\bf\bar{x}})$, where $a_{4}= B^2/A^2$.
The geometrical factor $a_{4}$ in front of the mass and source terms
is
redshifting away their contributions near the horizon, and can be
obtained by considering the Klein-Gordon equation in the given
spacetime and coordinates. Since the background metric is vacuum, the
conformally invariant wave equation is also trivially included in the
model.

Eq.~(\ref{eq:rw3d}) describes the time evolution of general linear
fields around non-rotating black holes, in the sense that the
evolution of all multipole perturbations can be encoded in this
function $\Phi(t,{\bf\bar{x}})$ of the four spacetime variables.
Initial data for $\Phi$ and $\dot{\Phi}$ can always be obtained by
referring back to Eq.~(\ref{eq:rw}), but can also be given
arbitrarily,
and then correspond to some, probably complicated, superposition of
decoupled perturbations.  This generality permits the direct study of
waves with different wavelengths and arbitrary angular structure.
Despite the vector notation in presenting Eq.~(\ref{eq:rw3d}) it
should be stressed that the field $\phi$ does not (with the exception
of the case $K_{1}=2M,K_{2}=0$), represent a physical scalar in
spacetime.  Our construction is, in the general case, completely
coordinate dependent, and the unit weight factors in
Eq.~(\ref{superpose}) are an arbitrary choice.  In the scalar field
case, general coordinate transformations of the underlying spacetime
(e.g., different time slicings, boosted coordinates) are of course
allowed.

To complete the specification of the initial value problem for
Eq.~(\ref{eq:rw}), we need to supplement Eq.~(\ref{eq:rw3d}) with
boundary
conditions.  With respect to the exterior boundary, we adopt the
approach of evolving only the development of the initial Cauchy
surface, hence we place our boundary far enough that it will not
causally affect the spacetime domain we are considering. For the
simulations performed here, the initial data have compact support,
which enlarges the domain not influenced by the outer boundary by
about one crossing time.

In isotropic coordinates the event horizon is described by the
coordinate surface $\bar{r}_{H}(x,y,z)=M/2$. The use of
\sch slicing implies that the lapse function has
a zero there. The effects of angular momentum and curvature are
rapidly damped (see for example~\cite{Chandrasekhar83})
and the effective black hole potential decays exponentially
in $r_{*}$. Imposing an ingoing wave type condition
\begin{equation}
Z_{,t} - Z_{,r^{*}} = 0 \; ,
\label{eq:bc1}
\end{equation}
at some inner timelike world-tube at radius $r_{*}^{B}$
is applicable for any equation of the form of~(\ref{eq:rw}).
The transformation of condition~(\ref{eq:bc1}) to the coordinates and 
variables of our model equation yields
\begin{equation}
\Phi_{,t} = \frac{B}{A^{3}\bar{r}}
( {\bf\bar{x}} \cdot \bar{\nabla} \Phi + \frac{B}{A} \Phi) \; .
\label{eq:bc2}
\end{equation}
Condition~(\ref{eq:bc2}) becomes
equivalent to a static horizon condition ($\Phi=0$) when applied in
the
limit $\bar{r}_{B} \rightarrow \bar{r}_{H}$, but is actually
extremely accurate when applied at a timelike world-tube that is
inside the decay width of the potential, as it falls from
its peak value at $\bar{r}_{P}\approx 2$. This fact has been both
extensively tested and used in the literature.

\subsection{The nature of the solutions}

Here we touch upon issues regarding the model equation~(\ref{eq:rw3d}), 
especially as they apply to our adaptive computations. 

The merger waveform, due to the final interactions of the 
black holes as they coalesce, will be an important component 
of the detected signal, which when detected 
can presumably reveal intricate details of the coalescence 
process. Even the broad features of the merger waveforms are 
uncertain at present (and may require techniques such as those 
we are developing and testing here to compute). 

As we have emphasized, the later stages
of binary black hole mergers lead to highly distorted black holes that
can be treated surprisingly well by perturbation theory. Even the
entire collision process has proved amenable to a perturbative
treatment in certain regimes. Through this large body of work, the
following picture has emerged: during the very late stages of a binary
black hole merger, the highly distorted black hole formed in the
process rings down to its Kerr form through a progression of regular,
damped oscillations, with the least damped quasinormal mode quickly
dominating the picture. The frequency (wavelength) of the slowest
damped mode is an important feature of the solution in the ring-down
phase. Fully nonlinear simulations of such distorted rotating black
holes have been performed, and the ($m=0$) quasi-normal modes are
found to be the dominant feature of the waveforms\cite{Brandt94c}.

For full 3D black hole coalescence, leading to a rapidly rotating
black hole, at late times the $\ell=m=2$ mode is expected to dominate,
having a much higher frequency than its Schwarzschild counterpart
(perhaps two to three times higher). Hence, for realistic 3D
simulations, the resolution requirements will be higher than what
would be indicated by a naive treatment of Schwarzschild
perturbations. A powerful approach to modeling the ringing phase of a
rotating black hole would be the use of equations based on curvature
perturbations. Recently, certain members of the Teukolsky family of
equations have been integrated numerically as a {2+1}
problem~\cite{Krivan96a,Krivan97a}.  As some aspects of the numerical
integration of those equations e.g., coordinate systems, higher
dimensional simulations, are now under investigation, we chose here to
base our model in the more thoroughly understood dynamics of potential
type equations.

The potential of Eq.~(\ref{eq:rw3d}) can be modified to allow the study
of a larger range of ringing wavelengths.  To this end, we have
performed a numerical study of the slowest damped modes of 
Eq.~(\ref{eq:rw3d}) for various angular multipoles, using 1D
versions of the equations, as such a catalog is convenient for
calibrating the three-dimensional code.  The behavior of the real and
imaginary parts of the QNM frequency are rather interesting.  We
mention for example that the slowest damped mode of the equation has a
real frequency that is a monotonic function of $K_1$, for any fixed
$K_2$.  The imaginary part (decay rate) has an oscillatory dependence
on $K_1$. Different values of $K_2$ tend to introduce an overall shift
on the dependence of $\omega$ on $K_{1}$.  Figure~\ref{fig:qnm}
provides a convenient calibration of the slowest damped mode
wavelength and decay rate.

\subsection{1-Dimensional Considerations}
\label{sec:1D}
In this work we use the structure of the QNM's as an important
diagnostic tool for the efficiency, flexibility and accuracy of the
AMR suite.  For convenient calibration we use evolutions computed with
one-dimensional codes, both in the well behaved {\em tortoise}
coordinate, which approaches $-\infty$ at the horizon, and the
isotropic radial coordinate, which goes to a finite value of $r=0.5$.
The tortoise coordinate is most naturally adapted to the problem,
accounting exactly for the infinite blue shift of a wave as it
approaches the horizon.  If an ingoing wave is resolved on an equally
spaced grid in the tortoise coordinate near the peak of the potential,
it will remain so as it propagates towards the horizon, even as its
wave is highly blue shifted physically.  The tortoise coordinate can
be considered as a natural adaptive mesh for the problem, and the 1D
code in this formulation can easily give accurate results for a given
$\ell-$mode.

On the other hand, the isotropic coordinate description provides
important clues for the minimal amount of resolution required for the
3D code, as it is directly related to the 3D Cartesian coordinates.
In this coordinate, an ingoing wave will be shifted to much shorter
wavelengths as it approaches the horizon, and hence the wave will very
rapidly become difficult to resolve as it propagates in.  This is an
effect of the Schwarzschild slicing we are using in the model
equation. This effect can be seen alternatively by
considering the reduction of the wave speed as one approaches the
horizon: with the inner boundary at an isotropic radius of $r=0.6$,
(Schwarzschild radius 2.016), the speed drops by thirty-fold to about
0.027.

In spite of these considerations, one might think that the treatment
of the equation near the horizon would not necessarily have
important effects on the waves propagating far from the black hole,
for several reasons.  First, waves should be essentially ingoing in
this area, second, any reflections due to poor treatment here will be
rapidly redshifted as they climb out of the black hole potential, and
finally,
the potential barrier in Fig.~\ref{fig:potential} acts to protect the
outside observer.  However, it turns out that significant fractions of
the emitted signal are influenced strongly by the inner region near
the horizon, i.e., $\bar{r}_{B} < \bar{r} < \bar{r}_{P}$.  In
particular, poor resolution of the inner region and inner boundary can
have very adverse effects on the final waveform as shown below.

We demonstrate these points with a series of 1D simulations for an
$\ell=2$ wave in Fig.~\ref{fig:new1dfigure}.  We compare three
simulations that illustrate the importance of proper treatment near
the horizon.  Initial data consist of a Gaussian packet localized to a
region outside the potential barrier with vanishing initial time
derivative.  The parameters in the potential correspond to the
standard Regge-Wheeler potential. The integration domain extends from
an isotropic radius of $r=0.6$ ($r_{*}=-7.56$), to $r=100$.  The
tortoise based evolution attains adequate accuracy for our purposes
already for resolution $N=200$, where $N$ is the number of grid points
($\Delta r_{*}=0.5$); for comparison purposes we use a more than
sufficient $N=2000$ grid. The solid line in both panels of
Fig.~\ref{fig:new1dfigure} shows the signal recorded at a radius
$r=10M$, as obtained with the tortoise code. After an initial burst of
waves, the quasinormal modes of the black hole dominate, and are well
resolved. We use this as our fiducial run against which we compare
other simulations.

We now examine results obtained with the isotropic radius, which gives
an indication of the problems that will be encountered in 3D. In
Fig.~\ref{fig:new1dfigure} the dashed lines show results obtained with
the isotropic code using a fixed resolution of $\Delta r = 0.05$
(upper panel) and $\Delta r = 0.02$ (lower panel), which corresponds
to $N=2000$ and $N=5000$ respectively.  The ingoing boundary condition
is always applied at the same radius, $r=0.6$.  For large $r$, the
tortoise and isotropic coordinates are very similar, and therefore
meager resolutions of the order of $~0.5$ would be adequate to resolve
the burst phase (the first part of the signal, which overlaps
perfectly).  But near the horizon, the wave is woefully under-resolved
in isotropic coordinates, even with a grid that is ten times finer
($\Delta r = 0.05$), which shows its wrath on the waveform at late
times, i.e., when the under-resolved region near the horizon makes
causal contact with the observer.  With increased resolution (lower
panel) there is marked improvement, and near convergence to the true
QNM signal.  In contrast the solution for $\Delta r = 0.05$ (upper
panel) illustrates the subtle effect on the phase and frequency of the
QNM, even for what would be thought of as a very fine resolution.  The
frequency shift in the QNM suggests that the poor resolution of the
horizon region {\em modifies the effective potential} experienced by
the gravitational wave.  This argument is made more plausible by the
fact that, even in flat space, a reflecting potential near the origin
supports a finite number of QN-modes, see~\cite{Gomez94b,Schmidt}.

The severity of the effect suggests we have a first rate test problem
on which to apply AMR techniques. In order to obtain a sufficiently
resolved simulation with a uniform grid we had to use more than $25$
times the basic tortoise resolution of $N=200$.  In this case, it is
possible in 1D, but in 3D such a factor would lead to $25^{3} N^{3}$!
We will turn into the handling of this behavior in the next section.

Those adverse effects of poor resolution around the horizon suggest
that this is an issue to be kept in mind in more general
simulations. In non-linear numerical relativity simulations the time
slicing is dynamic and generally has a finite lapse at the horizon.
Despite the reduced blueshift experienced by a wave (the most
demanding factor), other physical factors also compete for resolution:
(a) the geometrical volume factor (since we are using Cartesian
coordinates); (b) the need for sufficient sampling of the inner part
of the potential (see Fig.~\ref{fig:potential}) in order to accurately
capture the correct scattering and development of the quasinormal
modes; and (c) the need to guarantee convergence for any type of
analytic condition near an excised horizon area.
Hence, also in implementations of apparent horizon boundary conditions
(AHBC), one will need to be careful to ensure that the waves going in
at the horizon be properly resolved, and that the boundary condition
allows ingoing waves to propagate off the grid, if one is to get the
black hole dynamics correct.  To date, even in successful
implementations of AHBC, only the longitudinal part of the spacetime
has been tested.  These AHBC simulations need to be applied to
spacetimes with dynamic black holes for which waveforms can be
extracted, such as those recently studied in
Refs.\cite{Camarda97a,Camarda97b,Allen97a,Allen98a} to ensure that the
low amplitude wave are not adversely affected by the treatment of the
spacetime near the black hole.

\section{The Numerical Algorithm: Multi-Dimensional Hierarchical
Linked List AMR}
\label{sec:code}

Conventional adaptive mesh refinement assumes a data structure based
on a hierarchical tree of embedded refined sub-grids
(\cite{Berger84}).  In one dimension this hierarchy is simple and
efficient -- each sub-grid is just a spatial interval demarcated by an
upper and lower bound -- however in multi-dimensions one must first
demarcate regions of refinement into clusters and then for each
cluster a refined rectilinear sub-grid must be placed over it (for
example \cite{Berger91}). The quality by which one is able to match
the desired refinement topology with rectilinear boxes is given by an
input parameter -- the clustering efficiency -- which when set to
$100\%$ produces a minimal refinement topology.  However, depending on
the refinement topology itself, this may mean the production, storage
and management of may thousands of `little boxes'. This number of
boxes (and hence their associated computational cost of management)
can be reduced considerably by reducing the clustering efficiency but
this inevitably over refines the computational domain and thus
increasing once more the computational cost this time of integrating
the
solution. Therefore, usually a balance needs to be attained between
the increased computational overheads required to achieve and maintain
a grid hierarchy with a $100\%$ clustering efficiency, hence many
boxes but no over-refinement of the computational domain as compared
to a reduced clustering efficiency, hence less refined boxes but
over-refinement of the domain.

To avoid such scenarios we have instead adopted the AMR method
proposed by Wild and Schutz \cite{Wild98a}. Here they build mesh
refinement that precisely matches the desired topological
requirements, but without clustering algorithms.

Their method centers on the representation of mesh refinement by a
simple
hierarchy of refinement levels
\begin{equation}
{\cal R}_0\cup{\cal R}_1\cup{\cal R}_2\cup . .
{\cal R}_l . . \cup{\cal R}_{TotalLevels} \;\; ,
\end{equation}
where ${\cal R}$ is a uniform distribution of grid points of arbitrary
topology and the subscript indicates the magnitude of the spatial and
temporal resolution such that ${\cal R}_0$ has the least resolution
and ${\cal R}_{TotalLevels}$ the greatest.  For neighboring
refinement levels ${\cal R}_l$ and ${\cal R}_{l+1}$ the boundaries of
the former always contain those of the latter ({\em i.e.} ${\cal
R}_{l+1}\in{\cal R}_l$)  and their spatial {\em and} temporal
resolutions differ by an integer {\em refinement factor} $\eta$ we
choose to be two.

To construct each refinement level, in 3-D Cartesian co-ordinates,
three sets of one dimensional double linked lists are required (one
set along each of $x$, $y$ and $z$-dimensions.)  These lists are in
turn built from a data structure (called nodes) which contain: 
\begin{itemize} 
\item A hexahedron of grid points.
\item Six sibling pointers (two for each.
dimension) for the double links between neighboring nodes.
\item Eight offspring pointers to allow each grid point 
present to spawn a new,
offspring, node.
\item  One parental pointer to identify the node it is
refining.
\item A flag to indicate whether it requires refinement or not.
\end{itemize}

An example of this type of mesh topology construct is shown in
Figure~\ref{fig:3d_refinement_example}.  Here the coarse grid ${\cal
R}_0$ (large cubes) consists of a $4\times3\times2$ lattice of nodes
and therefore represents a $8\times6\times4$ lattice of grid points
(assuming a refinement factor $\eta=2$.)  The smaller nodes are
refinement level ${\cal R}_1$.  Here it is clear to see why eight are
required to refine a parental node.

To evolve the refinement hierarchy, we use a modified Berger-Oliger
algorithm
\cite{Wild98a}. Here the concept of
each refinement level being integrated by one time step and then only
being integrated again once all higher refinement levels have been
integrated to the same temporal point, remains unaltered.
However, whenever neighboring refinement levels ${\cal R}_l$ and
${\cal R}_{l+1}$ are temporally coincident then only the subset ${\cal
R}_l\cap{\cal R}_{l+1}$ of the ${\cal R}_l$ solution that can
numerically influence the boundary of ${\cal R}_{l+1}$ is replaced by
that in ${\cal R}_{l+1}$, as compared to the Berger-Oliger approach
which injects the complete ${\cal R}_{l+1}$ solution onto ${\cal
R}_l$. This much reduced inter-refinement communication can help
performance on parallel computing platforms in situations where the
domain decomposition strategy for distributing the mesh refinement's
topology across processors does not ensure regions of refinement are
aligned with their parental region. Thus, potentially expensive
inter-processor communications are reduced from that of a refinement
level's volume to just its surface area.

To integrate the solution on each refinement level the AMR method
(written in C) traverses all linked lists in the {\em x}-dimension
-- one node at a time -- by using the appropriate pointers.
For each node the necessary finite difference stencil is read
from memory and passed to a separate program (written in
Fortran90). Here, the solution for the new time level is
computed and then returned to the AMR code to be stored
(in this way AMR is independent from the simulation --
it choreographs the topology of the mesh and the storage of the
solution therein, nothing more.)

The mesh is modified periodically to ensure its topology satisfies
that of the evolving solution. This process is always carried out
from ${\cal R}_{TotalLevels}$ downwards, such that for each
refinement level nodes are checked against the refinement criteria
to determine whether offspring nodes need to be created or destroyed.

The AMR method is made parallel on shared memory architectures using a
simple one dimensional domain decomposition strategy whereby the total
number of lists in the direction of traversal are distributed evenly
between the available processors~\cite{Wild96}.  Although simple, such
a scheme should not be under-estimated since it has the advantage of
ensuring every processor has a $100\%$ work load during the
Berger-Oliger integration cycle (a
great advantage when the workload per grid point is high as is the
case when considering the fully relativistic Einstein equations.)
Additionally it introduces a minimal change to the code: one must now
ensure each processor maintains temporal synchronization during the
Berger-Oliger integration cycle~\cite{Berger84}.

The discretization of the model equation within the AMR framework is
straightforward and is done using a three-level leapfrog scheme.
Interior
points are updated with the regular nine-point stencil, which is for
the
centered differencing of first and second order derivatives. Boundary
points are masked and discretized with sidewise differences.

\section{3D Adapted Computations}
\subsection{Prescribed Mesh Refinement}
\label{sec:static}

It is clear that in a 3D integration of the model equation some sort
of adaptive grid is required at least in the region near the horizon.
We have as our guide the model 1D problem above, where we saw in
Sec.~\ref{sec:1D} how critical it was to resolve the region around the
horizon.  Equal spacing in the tortoise coordinate gives constant
resolution coverage of a blueshifting wave as it approaches the
horizon, which translates into increasing resolution requirements in
the isotropic coordinate, and hence in the Cartesian coordinates.
Indeed unigrid (i.e., single uniform spacing) 3D Cartesian evolutions
with up to $300^{3}$ grids covering a region from the horizon to a
distance $r \approx 100M$ produce woefully inadequate waveforms, as expected
from the 1D considerations above.  The question we address next is how
to provide the adaption.

We first investigate whether one can use certain {\em
prior knowledge} of a system to construct an adapted predefined mesh
appropriate to the problem.  We hence prescribe successive refinement
layers around the horizon, whose topology and location remains
constant throughout the evolution.  This approach is a powerful and
efficient use of the AMR infrastructure in three dimensions, which is
easily combined with dynamic refinement in other parts of the domain.

Here we illustrate this technique with an example of seven successive
layers~(Table \ref{table:grids}) of refinement around the black hole
horizon, which were motivated by studies of the 1D equation in isotropic
coordinates. A valuable guide is the study of two radially ingoing 
null geodesics $\bar{r}_{1}(t)$, $\bar{r}_{2}(t)$, and correspondingly 
the allocation of resolution so as to roughly preserve the number 
of grid points in-between the geodesics as a function of radious.
To test this nested grid structure, we set up an initially time symmetric
quadrupole Gaussian pulse in the Regge-Wheeler function $\Phi$ located
at $10M$. Specifically, the initial data takes the form:
\begin{eqnarray}
\Phi & = & e^{-\kappa(\bar{r}-\bar{r}_{c})^{2}} R(\theta,\phi) \; , \\
\dot{\Phi} & = & 0 \; ,
\end{eqnarray}
where $\kappa$ and $\bar{r}_{c}$ control the width and location of
the shell, and $ R(\theta,\phi) $ sets an appropriate angular
dependence,
which for our standard comparison with the 1D Regge-Wheeler equation
we adopt to be a quadrupole:
\begin{equation}
R(\theta,\phi) = \sin^{2}{\theta} \cos{(2 \phi + \pi/2)}
\end{equation}
(the rotated form of $R$ in this particular case is explained in
the Appendix).

The accuracy in resolving the QNM structure using such nested
refinement levels is demonstrated in
Figs.~\ref{fig:famazing},\ref{fig:amazing}.  Fig.~\ref{fig:famazing}
shows results in linear scales, showing the familiar ringing waveform,
while Fig.~\ref{fig:amazing} shows the same simulation in a
logarithmic scale, to bring out different aspects of the results.  The
comparison is done at the peak of the potential (similar signals are
obtained e.g., at $r=10M$).  The base grid is at $\Delta x = \Delta y
= \Delta z =0.5M$ resolution and extends to $r = 50M$.

The solid lines are 1D results, using about $N=1000$ radial points
using the {\em tortoise code}.  The 3D solution converges to the
correct (1D) one as more layers of refinement are added (the coarse
grid evolution is too erroneous to display).  Using three refinement
layers (finest level at $\Delta x = 0.0625M$, dashed line)
considerably reduces the reflections, but still has a serious effect
on most of the QNM structure.  Further refinement (dotted line) to
seven levels (0.0039062M) captures the correct QNM frequency and decay
rate, over {\em orders of magnitude} in the decay. Further {\em
enlargement} of the thickness of the layers reduces the amplitude loss
at the expense, of course, of computational time.

These results have been obtained with refinement levels using {\em
nested spheres}, which naturally adapt to the geometry of a central
black hole.  Even though the underlying coordinate system being used
is Cartesian, by using {\em linked lists} this structure is still
straightforward and natural to create.  With more traditional grid
structures, if one attempted nested refinement levels it might be
easier to create {\em nested cubes}.  With the linked lists we can
also easily create such refinement structures and study their effect
on the results.  Using this nested box construction, we have
reproduced results for various refinement layers.  The results are
excellent in both cases, but there is a slight improvement in the
quality of the modes when using boxes with width equal to the diameter
of the corresponding sphere.  This is expected, given the considerably
larger refinement volume introduced by nested cubes.  This result
suggests that using linked lists to construct spheres within spheres
one can match the accuracy of the nested cube waveforms but using
$\pi/6$ of the memory.  This non-negligible saving is inherent to the
linked list approach, not to AMR, and could equally well be applied to
the construction of uniform grid codes.  Furthermore, in addition to
handling irregular outer boundaries linked lists are highly amenable
to the construction of an internal boundary, e.g., within a black
hole's horizon as shown here.

Longer evolutions of the same initial data reveal an interesting
phenomenon associated with the presence of different refinement
levels. Observers inside a given refinement level will observe
numerical reflections arising at the refinement level boundary
with the next coarsest grid. In our simulations those reflections
were well below the interesting signal levels, but this may
change in non-linear simulations with more meager resolutions,
or refinement factors larger than two.

To summarize the results of this section, we note that the essential
dynamics of a black hole in the linear regime was resolved with a
fixed, refined mesh structure.  In this approach we use knowledge of
the system to place appropriate refinement levels where they will be
needed.  We also found that the underlying geometry of the refinement
levels can be flexible, i.e., we found that nested boxes and nested
spheres both do an excellent job of resolving the calculation,
although the nested spheres allow considerable memory savings.  We
note, however, that fixed refinement must be used with care: if highly
resolved waves are allowed to propagate into regions that are unable
to resolve a wavelength, spurious reflections. (See
Wild~\cite{Wild96} or Wild and Schutz\cite{Wild98a}
for a detailed analysis of reflection and transmission properties of
waves crossing various grid interfaces, along with strategies to
handle such problems).  Extending the refined domain as the generated
signal propagates outwards, away from the highly resolved horizon
region, is a natural application for {\em dynamic} mesh refinement, to
which we turn next.

\subsection{Dynamic Mesh Refinement}
\label{sec:dynamic}

Using our model equation we also explored the issue of outgoing waves
-- with wavelengths typical of black hole QNM's -- propagating
outwards on an initially coarse grid, which would not provide adequate
resolution if not refined.  Depending on the initial burst of waves
hitting the black hole, different outgoing radiation patterns and
wavelengths may emerge, which cannot be predicted ahead of time.  We
now examine approaches for supplying the necessary resolution to these
waves, which must be tracked and resolved as they propagate.  In what
follows we maintain the static refinement necessary to resolve the
region around the horizon (including all seven levels listed in Table
\ref{table:grids}), and explore different methods to track the waves
as they propagate to large distances from the black hole.  In this
``hybrid'' refinement strategy we allow the adapted grids to follow
the waves where ever they go, both away from the hole and into the
highly refined region near the hole.

A crucial aspect of AMR is the choice of refinement criteria. A
common {\em general purpose} estimator for dynamic refinement is the
intrinsic truncation error estimate. Such an estimate of the
truncation error of a numerical solution, $\tau$, can be defined as:
\begin{equation}
\label{eq:trunc_est}
\tau \propto |{\cal Q}^2_{\Delta s,\Delta t}\phi - {\cal Q}_{2\Delta
s,2\Delta t}\phi| \; ,
\end{equation}
where ${\cal Q}$ is the finite difference evolution operator used to
integrate the solution and $\Delta s$, $\Delta t$ are the spatial and
temporal resolutions, respectively. Thus, the truncation error is
computed by taking the difference of the solution obtained using two
regular time steps of $\Delta t$ for a spatial resolution $\Delta x$
from the solution obtained using just one large time step of $2\Delta
t$ on a coarsened grid of $2\Delta x$. Thus for a given point in the
computational domain if the magnitude of Eq.~(\ref{eq:trunc_est}) is
larger than some specified amount (e.g. $\tau \geq threshold$) then
locally that region of the computational domain is refined (see
\cite{Berger84}.)

Such criteria in lower dimensional work have been demonstrated to work
very well in demanding circumstances (e.g.
\cite{Choptuik93,Masso94b,Wild96}.)  However, applying the truncation error
estimator Eq.~(\ref{eq:trunc_est}) as the refinement criteria, verbatim,
can fail.

To understand what we mean by ``fail'', consider the {\em radial
scaling} of the waves as they propagate away from the black hole and
the impact of this effect on the refinement criteria.  The perturbed
metric variables have strong radial falloff behavior (for
sufficiently large radii $r$, $h\sim 1/r$).  In our single wave
equation, this is very clear, although in a more general case, as one
would encounter in the full set of Einstein equations, there will be a
mixture of different falloff rates depending on the variables, the
gauge, and so forth.  But physically the wave part of the solution,
given by $h$, will decay as it propagates away.  Since refinement
criteria like Eq.~(\ref{eq:trunc_est}) are proportional to the local
magnitude of the solution, the decaying nature on the function
being refined undermines the effectiveness of the refinement
criteria's ability to place refinement.

\subsubsection{Adaptive refinement studies for a pure $\ell=2$ mode}

We illustrate many of these issues by considering the effect of
different refinement criteria on a 3D simulation for a pure $\ell=2$
wave, as discussed in the section above.  We consider a general packet
containing many $\ell-$modes, taking advantage of our generalized wave
equation, in the next section.  We note that these simulations are
much more demanding than those discussed in the section above, where
waveforms were measured near the peak of the potential near the
horizon.  In this section we study the propagation of waves far from
the black hole, and our ability to dynamically track and resolve them.

Consider the refinement criteria Eq.~(\ref{eq:trunc_est}) with the
demand that $\tau\leq0.002$ throughout a computation domain which
extends to $300M$ and a base resolution of $2M$.  In this simulation
of the pure $\ell=2$ mode, we allow one level of refinement.  In
Fig.~\ref{fig:growth} we show the growth number of grid points for
several adaptive mesh refinement simulations.  As the waves propagate
out from the black hole, they of course sweep out an ever larger
volume.  Plotting the percentage growth of dynamic grid points,
relative to the number of initial points, as a function of time (see
dotted line in Fig.~\ref{fig:growth}) we see that the growth of the
dynamic refinement occurred only over a finite period of time,
eventually evaporated even as the waves expand to larger volumes!
This is due to the decay of the wave amplitude below the level at
which it can be captured by the specified truncation error limit.
This is contrary to our AMR requirements for capturing and
transporting the physical waveform to some arbitrarily distant
detector.  The solid and dashed lines correspond to other refinement
criteria discussed below.

Indeed the net effect of such evaporation on a waveform is to
dramatically perturb it, since waves once initially captured within
some region of mesh refinement will not remain contained and instead
``leak'' back across the dynamic refinement boundaries onto coarser
resolutions.  To demonstrate this behavior, in
Fig.~\ref{fig:tracking1} we show the actual waveform obtained for the
same simulation, again demanding that $\tau\leq0.002$ throughout the
computation domain.  Initially this results in the first two out-going
modes being contained within the dynamic refinement, although with
only one level of refinement the wave is still not resolved enough.
However, by the time these modes have passed a detector placed at
$125M$ only the first mode remains captured within the refinement --
the second mode now trails in its wake.  Moreover, because the
evaporation of the grid is non-smooth (see Fig.~\ref{fig:growth}
dotted line), this has an effect of perturbing the waveform which
becomes highly inaccurate.  Here the solid line is the correct
waveform obtained from 1D simulations in the tortoise coordinate.  The
dotted line is the truncation-error based AMR result, clearly
revealing the deterioration of the second mode which now trails the
dynamic mesh refinement.  (For a discussion on the effects moving
boundaries have on perturbing waves straddling mesh refinement see
\cite{Wild96}.)  One cannot cure this effect by simply tightening the
tolerance on the refinement criterion since this can only improve things for
awhile, but eventually as wave decays away, the same effects will
be seen.

Such behavior should not be thought of as indicative of truncation
error estimators. This dynamic refinement behavior is generic to
refinement criteria which do not take into consideration the
underlying nature of the solution's falloff. For example, employing a
refinement criterion based on a {\em norm} of the solution:
\begin{equation}
N_{1}= c_{1} |\Phi|^2 + c_{2} |\Phi_{,t}|^2 \; ,
\label{eq:norm}
\end{equation}
produces similar effects, as shown by the dashed lines in
Fig.~\ref{fig:growth} and Fig.~\ref{fig:tracking1}.  In
Fig.~\ref{fig:growth} we see the initial growth and final decay of
dynamic grid points, and in Fig.~\ref{fig:tracking1} we see a similarly poor
waveform.  Interestingly however, the evaporation of the dynamic
refinement created in this case is more regular and hence shorter
lived than that for refinement criteria based on estimating the
truncation error.  This difference can be traced to the fact that we
found the truncation error estimator Eq.~(\ref{eq:trunc_est}) produced
a ``noisier'' estimate of where refinement was required and was therefore
less able to
create a regular pattern of refinement than the norm based criterion.

To overcome this grid evaporation requires an appropriate scaling the
refinement criteria. For example, for the norm refinement criteria by
scaling it radially with $1/r^2$ results in both the first and second
outgoing modes remaining captured within the mesh refinement -- in
principle forever.  (The effects of dispersion will eventually
negate this effect however.) This is shown by the dotted line in
Fig.~\ref{fig:tracking2}. Here, the trailing part of the signal that
falls outside the dynamic region of refinement has deteriorated
considerably. The quality of the signal can be increased further by
introducing a second level of dynamic mesh refinement as shown by the
dashed line in Fig.~\ref{fig:tracking2}.  We obtained similar results
by suitably scaling the truncation error criterion, although as
before, with a slightly noisier solution.

Finally, for simulations of a pure $\ell=2$ multipole wave hitting a
black hole, we show the evolution of the grid refinement structure as
the wave propagates out. Again we use a base grid extending out to 300M,
with a resolution of 2M. Two levels of dynamic AMR track outgoing waves
using the norm based refinement criteria (Eq.~(\ref{eq:norm})) with the
correct radial scaling of $1/r^2$. Seven layers of prescribed refinement
resolve the region near the black hole horizon.

Fig.~\ref{fig:mesh1} shows the mesh structure and isosurfaces for
the outgoing pulse at time $t=100M$, and Fig.~\ref{fig:mesh2} show the
same system at time $t=200M$. The boundaries of the refinement
layers are indicated by the zig-zag lines. Two such layers engulf the
outgoing burst and of the seven layers resolving the hole, only three are
shown.

At the $t=100M$ stage (Fig.~\ref{fig:mesh1}) the dynamically prescribed
layers tracking the outgoing burst still
overlap with the prescribed layers around the hole. The two wavelengths
(depicted as isosurface shells) are captured by the ${\cal R}_{2}$ refinement
and will be propagated accurately outwards.  The third oscillation
resides just outside ${\cal R}_{2}$, a fact which will prove
detrimental to its later accuracy. At $t=200M$ (Fig.~\ref{fig:mesh2}) the
boundaries of the refinement layers covering
the hole and the outgoing signal are clearly separated.
The resulting coarse region in-between cannot support the lower amplitude
trailing signal which consequently becomes heavily distorted.

In such simulations, with expanding wavefronts, the price one
must now pay for keeping a solution correctly captured within 3D
dynamic refinement is to inevitably have to increase the number of
grid points within the computational domain.  This is shown by the
solid line in Fig.~\ref{fig:growth}.  This rapid growth of the
consumed resources, as the wavefront expands onto the coarse grid,
will be typical of the more general black hole coalescence problems.
Thus, for the future effective use of AMR in black hole simulation,
one must address directly the question of what amount of the signal
needs to be captured with high resolution grids and set the refinement
criteria accordingly (appropriately scaled by $r$) to distribute the
refined grids around the strongest part of the signal, i.e., the
initial burst, and the largest QNM oscillations immediately following.

\subsubsection{Simulations of general pulse hitting a black hole}

The previous discussion was based on a single initial data set, which
contained only a pure $\ell = 2$ angular structure. We turn now to a
more general wave packet, probing the AMR performance
on more complex wave patterns that could be encountered in the
general black hole coalescence problem. To this end, we consider
general initial data, for example data that have compact support in a
three dimensional volume which does not surround the black hole, for
example
\begin{eqnarray}
\Phi & = &
e^{-\kappa((\bar{x}-\bar{x}_{c})^{2}+(\bar{y}-\bar{y}_{c})^{2}
+(\bar{z}-\bar{z}_{c})^{2})}  \; , 
\\
\dot{\Phi} & = & 0 \;  .
\end{eqnarray}

It is obvious that such evolutions would be intrinsically
three-dimensional. The emitted signal has a distinct ``burst'' phase,
containing radiation of considerably high frequency.  The presence of
high harmonics, with the corresponding short wavelengths makes the
accurate evolution of such data more demanding than simpler
superpositions of low lying modes. Here we illustrate the ability of
the AMR suite to capture complex solution patterns. The refinement
criterion used was norm based (Eq.~(\ref{eq:norm})) and one level of
dynamic refinement.

Fig.~\ref{fig:glory} shows a planar slice of mesh structure and isosurfaces of
dumb-bell shaped initial data, after 200M of evolution.
This ``burst shell'' of overlapping high-frequencies will eventually
be succeeded by the more regular pattern of QNM ringing. Here, these high
frequency features have no possibility of being accurately transported to a
distant detector without dynamic AMR.

The corresponding three dimensional mesh structure at t=200M is shown in
Fig.~\ref{fig:3dmesh}. Here octant symmetry, in connection with the initial
data of compact support, produces ``voids'' in between the out-going
wavefronts. This corresponds to the delayed arrival of the ``mirror'' data in
the computational domain. In these regions dynamic refinement senses the
absence of a strong signal and therefore leaves unrefined.

\section{Conclusions and Future Directions}
\label{sec:summary}

This work presents an introduction to the potential and problems of
full 3D adaptive mesh refinement in numerical relativity.  For the
purpose of our investigation in adaptive three dimensional
computations, we introduced a general class of three-dimensional
partial differential equations that capture important aspects of black
hole interactions. For select parameters those equations correspond
rather directly to physical models of scalar, electromagnetic and
gravitational interactions of black holes, hence physical processes
can be described (in the weak limit) in terms of those
equations. Preliminary analysis of those equations outlines aspects of
their behavior, in particular the dependence of the ring-down signal
on the parameters, and the resolution requirements of the solution in
the near horizon region in isotropic static slicings of the black
hole.

We presented a series of {\em fully resolved} three-dimensional
computations involving dynamics in black hole spacetimes. The linear
nature of the problem does not reduce the large dynamic range of the
black hole potential, which manifests itself through the strong radial
dependence of the coefficients of the equation.  Our adopted gauge
accentuates the large resolution requirements near the
horizon. Prescribed fixed refinement was successfully used to provide
the required resolution in that region.

Dynamic refinement was used to propagate signals into the exterior
domain.  The necessary {\em scaling} of refinement criteria by
appropriate power of $r$ was discussed, along with the impressive
growth of grid points occurring when the outgoing burst is resolved
dynamically as it propagates. The idea of {\em norm-based refinement}
was introduced, which works well in conjunction with a selective
tracking of the strongest part of the signal. Our considerations are
based of course on our model equations, but may be useful for the
Einstein problem. In particular we propose the (appropriately scaled)
norm $N_{2} = c_{1} |\Psi_{4}|^2 + c_{2} |\Psi_{4,t}|^2$ where
$\Psi_{4}$ is a locally computed component of the Weyl tensor
describing, in vacuum, outgoing radiation.  This might prove to be an
effective outer refinement criterion, and should be tested on model
problems
involving the full set of Einstein equations.

We are presently extending this work in several directions.  Our study
directly demonstrates that 3D investigations of black hole physics in
the linearized limit can directly benefit from AMR methods.
Extensions of the presently described models involving rotation and
different stationary coordinate systems are
underway~\cite{Papadapoulos98b}.
Studies of non-linear systems are also underway (e.g., in connection
with ADM evolution of single black hole spacetimes in
three-dimensions.) The considerable complexity of the adaptive mesh
infrastructure suggests comparisons of the HHL data structure with the
DAGH data structure, developed for the BBHGC alliance program, in
particular with respect to performance in three-dimensional
computations on parallel architectures.

\section*{Acknowledgments}

This work has been supported by the Albert Einstein Institute (AEI).
Calculations were performed at AEI on an SGI/Cray Origin 2000
supercomputer.  We have benefited significantly from interactions with
many colleagues at AEI and NCSA, especially Miguel Alcubierre,
Gabrielle Allen, Bernd Br{\"u}gmann, Randy Leveque, Joan Mass\'o,
Bernard Schutz, John Shalf, Bernd Schmidt, Wai-Mo Suen, and Paul Walker.
The work presented was substantially complete when we became aware of
a recent preprint \cite{Rezzolla97a} dealing with similar issues, but
without adaptive mesh refinement.  We acknowledge Andrew Abrahams and
Luciano Rezzolla for illuminating discussions.

\appendix

\section*{Numerical angular mode mixing and instabilities}

In this appendix we elaborate on a
feature of some of the equations~(\ref{eq:rw3d}), which is important
for stable long-term evolutions, a very desirable characteristic for a
numerical problem.

The PDE~(\ref{eq:rw3d}) is equivalent to an $\ell$-sequence of
separated one-dimensional PDE's.  It is of some importance to note
that from the point of view of {\em approximate} solutions to the
initial value problem the equivalence may break down.  By this we mean
that a discrete approximation to (\ref{eq:rw3d}) will not (in general)
be a {\em separable difference equation}, and hence numerical mixing
of angular modes is possible, and is generally the case unless some
techniques are devised to prevent it.

Some of the equations~(\ref{eq:rw3d}) demonstrate in the late stages
of numerical evolution a dramatic manifestation of such mode mixing.
For values $K_{1}=-6,K_{2}=0$, the equation represents axial
gravitational
perturbations and hence should exclude non-radiative solutions with
spherical or dipole symmetry ($\ell$=0,1). It turns out that an {\em
approximate} integration of the initial value problem is unstable with
respect to the $\ell =0$ mode, i.e., even if the mode is absent in the
initial data it will appear in the solution and it will exhibit
unbounded growth.

The origin of the unstable $\ell=0$ behavior can be easily seen by
inspection of the separated equation~(\ref{eq:rw}) for $\ell=0$.  The
effective potential is negative in the entire domain, hence a simple
examination of the dispersion relation ($\omega^2=\kappa^2-V$) for
spherical waves reveals the presence of local modes that are
exponentially growing.  Depending on the accuracy of the integration,
the manifestation of the instability may be delayed, but in finite
precision arithmetic it is bound to occur due to round-off error. In
our simulations it occurs typically after 100M of evolution for
$128^{3}$ base-grids. It manifests itself much earlier if a spherical
component is analytically introduced at the initial time.

For long time evolutions, with potentials admitting growing modes,
excluding the unstable $\ell=0$ mode is possible with the appropriate
use of boundary conditions. To this end, we restrict the integration
domain to an octant, and select to impose at least one condition of
{\em anti-reflection} across the planes defining the octant domain.
Such conditions eliminate spherical modes {\em even} for the
discretized equations.  The spherical nature of the background
conveniently allows one to rotate the coordinate system (this is the
of course the reason why multipoles of same $\ell$ but different $m$
values obey the same equations for non-rotating black holes). 

As an example, the $\ell=2,m=0$ mode with angular dependence proportional to
$3\cos(\theta)^2 -1$, will be decomposed, in a rotated frame $\hat{\theta}
=\theta + \pi/2$, to sum of quadrupole terms, of which the 
$\ell=2,m=2$ mode, with a further $\phi$ rotation by $\pi/4$ gives
$\sin(\hat{\theta})^2 \cos(2 \hat{\phi} + \pi/2)$ . This angular dependence 
admits antireflection conditions across the $x=0$ and $y=0$ planes.
Indeed, with the use of such conditions, the numerical integrations
show no sign of unstable growth for at least 500M of evolution time.

Axial perturbations will be present in full non-linear simulations,
and their dynamics will be governed, at least in some weak regime, by
a collection of coupled linear equations.  It is not clear that such a
linearized system of Einstein equations should have an unstable
spherical mode of the same type exhibited by our model PDE. This issue
warrants some more investigation. If this were the case though,
eliminating the instability would be very difficult for integrations
in the full spatial domain and non-spherical holes.

\bibliographystyle{prsty}
\bibliography{references}

\pagebreak

\begin{table}
\caption{\label{table:grids}
The first column refers to the refinement level, which is bounded
in the outside by the radial value of the second column. The
resolution
of each level (refinement by factors of two) is given in the third
column. The innermost grid point at which a given level approximates
the inner boundary is shown in the fourth column.}
\begin{tabular}{|c|c|c|c|}
\hline
Refinement level & Outer Bound (M)
& Resolution (M) & Inner Bound \\
\hline \hline
$r_1$  & 10.0  &  0.25      &  0.75       \\
$r_2$  & 5.0   &  0.125     &  0.625      \\
$r_3$  & 2.5   &  0.0625    &  0.625      \\
$r_4$  & 1.5   &  0.03125   &  0.625      \\
$r_5$  & 1.0   &  0.015625  &  0.609375   \\
$r_6$  & 0.75  &  0.0078125 &  0.6015625  \\
$r_7$  & 0.62  &  0.0039860 &  0.6015625  \\
\hline
\end{tabular}
\end{table}

\pagebreak


\begin{figure}
\caption{The effective potential $a_{3}$ for axial perturbations
$(K_{1} = - 6M, K_{2} = 0)$. The interesting range spans three decades in the
logarithmic scale. The inner region harbors a one way (ingoing) membrane
(the event horizon) which requires adequate resolution in order
to prevent reflections from the inner boundary and poor
sampling of the rapidly decaying potential. The middle region
determines essential features of the outgoing radiation (amplitude,
frequency, decay rate). The exterior region sees the wave
gradually transform into a radially propagating pulse with a fixed
wavelength.}
\label{fig:potential}
\end{figure}


\begin{figure}
\caption{Graphical representation of the frequency and decay rate
of the slowest damped mode for potentials of the type $a_{3}$.
Here $l=2$ and the parameter $K_{2}$ has been set to zero. The derived
values for $K_{1}=-6M,0,2M$ agree well with values tabulated in the
literature.}
\label{fig:qnm}
\end{figure}


\begin{figure}
\caption{We show 1D results for a gaussian wave packet hitting the
black hole Regge-Wheeler potential.  The solid line in both panels
shows the waveform measured at $r=10M$, obtained with an equally spaced
grid (in tortoise coordinate) of $\Delta r_{*} = 0.05$, or $N=2000$.
The dashed lines show results obtained with the isotropic code using a
fixed resolution of $\Delta r = 0.05$ (upper panel) and $\Delta r =
0.02$ (lower panel), which corresponds to $N=2000$ and $N=5000$
respectively.  The ingoing boundary condition is always applied at the
same radius, $r=0.6$.  For large $r$, the tortoise and isotropic
coordinates are very similar, and therefore meager resolutions of the
order of $~0.5$ would be adequate to resolve the burst phase (the
first part of the signal, which overlaps perfectly).  But near the
horizon, the wave is woefully under-resolved in isotropic coordinates,
even with a grid that is ten times finer ($\Delta r = 0.05$), which
shows its wrath on the waveform at late times, i.e., when the
under-resolved region near the horizon makes causal contact with the
observer.  With increased resolution (lower panel) there is marked
improvement, and eventual convergence of the QNM signal.}
\label{fig:new1dfigure}
\end{figure}


\begin{figure}
\caption{3-D mesh refinement using nodes. The large cubes (nodes)
represent a $8\times6\times4$ lattice of grid points that make up the base
grid ${\cal R}_0$. The smaller nodes are the refinement level ${\cal
R}_1$.}
\label{fig:3d_refinement_example}
\end{figure}


\begin{figure}
\caption{We show waveforms obtained for different levels of fixed
refinement in the interior region near the horizon, and its effects on
the quasi-normal mode structure.  The solid line is the 1D result,
which is progressively attained using three refinement layers
(dashed line) and seven refinement layers (dotted line).  Poor
resolution leads to under-sampling of the inner part of the potential
which combined with reflections from the ingoing boundary condition
generates an effective numerical potential with different frequency
and decay rate.  Fig.~\ref{fig:amazing} illustrates this point more
clearly with the use of logarithmic scale for $\Phi$.}
\label{fig:famazing}
\end{figure}


\begin{figure}
\caption{Logarithmic plot illustrating accuracy issues related to the
generation of QNM signals.  The time signal is extracted around 3M and
is compared with the corresponding one-dimensional result (solid
line).  The base grid has a resolution of 0.5M, which is too coarse to
resolve the black hole potential and hence produces a large reflected
wave (not shown).  Using three refinement layers (finest level at
0.0625M, dashed line) considerably reduces the reflections but
cannot reproduce most of the QNM structure.  Further refinement (dotted
line) to seven levels (0.0039062M) captures the correct QNM frequency
and decay rate over the whole domain of interest.}
\label{fig:amazing}
\end{figure}


\begin{figure}
\caption{Using the truncation error estimator Eq.~(\ref{eq:trunc_est})
without radial scaling reveals that the refined region ``evaporates''
as it propagates outward (dotted line).  This behavior is also true
with the norm refinement criteria Eq.~(\ref{eq:norm}) (dashed line).
However, appropriately scaling the norm refinement criteria
produces a continuous growth of dynamic gridpoints (solid line.)}
\label{fig:growth}
\end{figure}


\begin{figure}
\caption{Tracking of outgoing waves using one level of dynamic
refinement.  The signal as seen by an equatorial observer located at
125M. The base grid resolution is 2M. Using truncation error
refinement criteria (Eq.~(\ref{eq:trunc_est})) with no radial scaling
produces a poor quality signal (dotted line) compared to the the 1D
result (solid line).  The trailing part of the signal falls outside
the AMR captured region and deteriorates considerably.  This is also
the case for the norm refinement criteria (Eq.~\ref{eq:norm}) with no
radial scaling, shown as a dashed line.}
\label{fig:tracking1}
\end{figure}


\begin{figure}
\caption{Tracking of outgoing waves using one and two levels of
dynamic refinement with a ``scaled'' refinement criterion.  The signal
shown is seen by an equatorial observer located at 125M. The base grid
resolution is 2M. Using norm refinement criteria (Eq.~\ref{eq:norm})
with $1/r^2$ radial scaling captures and contains the first two
outgoing modes (dotted line) compared to the the 1D result (solid
line).  The trailing part of the signal falls outside the AMR captured
region and has inevitably deteriorated.  The quality of the signal
improves even more when two levels of dynamic refinement are used
(dashed line).}
\label{fig:tracking2}
\end{figure}


\begin{figure}
\caption{Mesh structure and isosurfaces for an outgoing pulse,
at time 100M. The grid extends out to 300M, with a base resolution of
2M. The boundaries of the refinement layers are indicated by the
zig-zag lines. Two such layers engulf the outgoing burst. Several
layers cover the hole, of which only three are shown. At this stage
the dynamically prescribed layers tracking the outgoing burst still
overlap with the prescribed layers around the hole. Two wavelengths
(depicted as isosurface shells) are captured by the $L_{2}$ refinement
and will be propagated accurately outwards.  A third oscillation
is seen to be just outside $L_{2}$, a fact which will prove
detrimental to its shape.}
\label{fig:mesh1}
\end{figure}


\begin{figure}
\caption{Mesh structure and isosurfaces for outgoing pulse,
at time 200M. The boundaries of the refinement layers covering
the hole and the outgoing signal are clearly separated.
The coarse region in-between cannot support the lower amplitude
trailing signal, which will become heavily distorted.}
\label{fig:mesh2}
\end{figure}


\begin{figure}
\caption{Planar slice of mesh structure and isosurfaces of
dumb-bell shaped initial data, after 200M of evolution.
A shell of overlapping high-frequency bursts will eventually
be succeeded by the more regular pattern of QNM ringing.}
\label{fig:glory}
\end{figure}


\begin{figure}
\caption{Three dimensional mesh structure. The octant symmetry
in connection with initial data of compact support produces
``voids'' in between the fronts, which correspond to the delayed
arrival of the ``mirror'' data in the computational domain. The
dynamic refinement senses the absence of strong signal and economizes
the grid.}
\label{fig:3dmesh}
\end{figure}

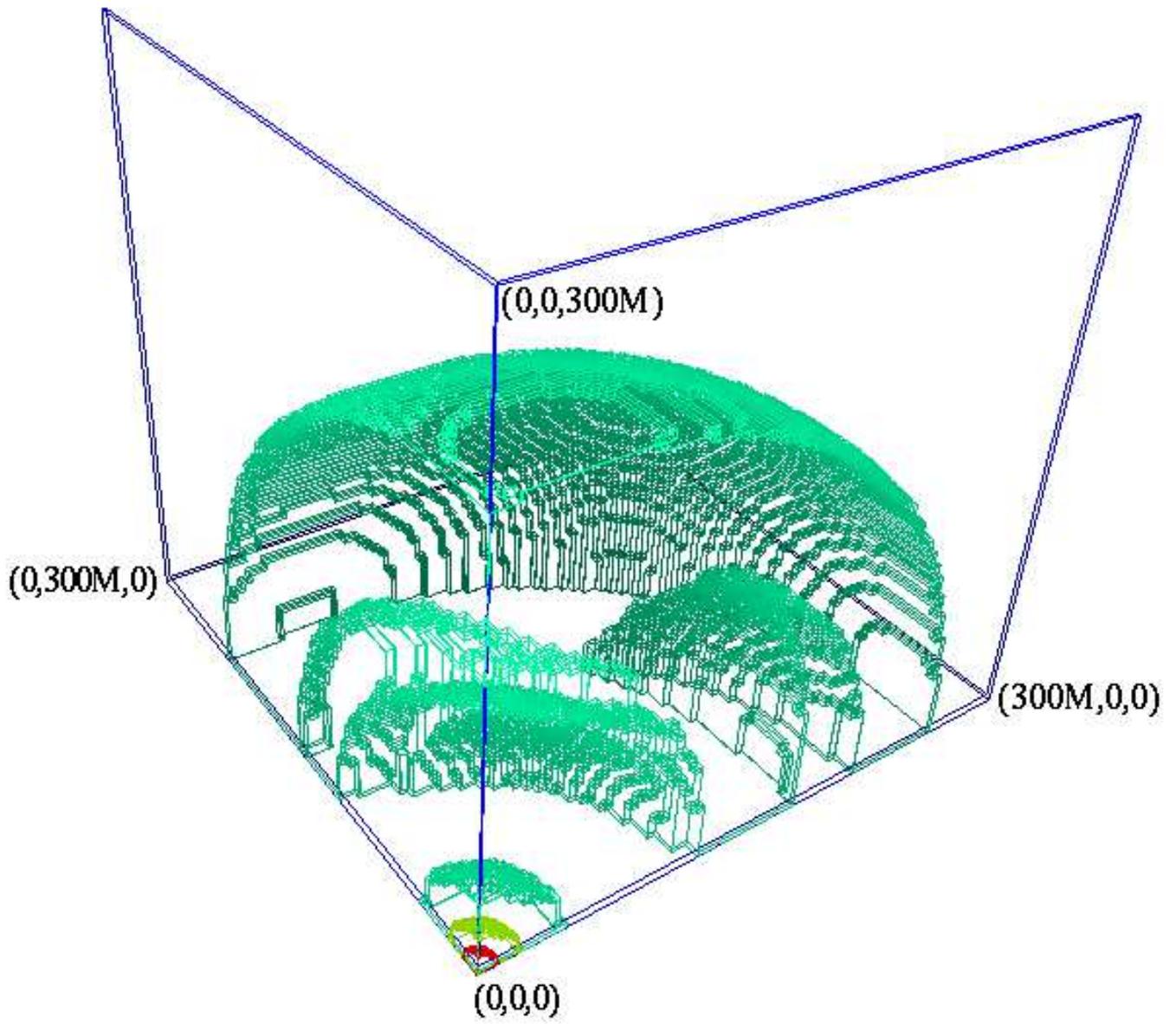
\end{document}